\documentclass[runningheads]{llncs}
\usepackage[T1]{fontenc}
\usepackage{graphicx}

\usepackage[utf8]{inputenc} 
\usepackage{hyperref}       
\usepackage{url}            
\usepackage{booktabs}       
\usepackage{amsfonts}       
\usepackage{nicefrac}       
\usepackage{microtype}      
\usepackage{xcolor}         

\usepackage{tikz}
\usepackage{amsmath,physics}
\usetikzlibrary{shapes}
\usepackage{soul}
\usepackage{url}
\usepackage{amsmath, amssymb}
\usepackage{enumitem}
\usepackage{multirow}
\usepackage{enumitem}
\usepackage{wrapfig}
\usepackage[font=small]{caption}
\usepackage[english]{babel}
\usepackage{algorithm}
\usepackage{algpseudocode}
\usepackage{subcaption}
\usepackage{multicol}
\usepackage{xspace}
\usepackage{tabularx}
\usepackage{ulem}

\newcommand{\subscript}[2]{$#1 _ #2$}
\newcommand{\pLAP}{\ensuremath{\pi_{\mathsf{LAP}}}\xspace}
\begin{document}
\title{Privacy-Preserving Fair Item Ranking}
\titlerunning{Privacy-Preserving Fair Item Ranking}
%

 \author{Jia Ao Sun\inst{1,2} \and
 Sikha Pentyala\inst{1,3} \and
 Martine De Cock\inst{3,4} \and
 Golnoosh Farnadi\inst{1,2,5}}

 \authorrunning{J. Sun et al.}

 \institute{Mila - Quebec AI Institute, Montréal QC Canada \and
 Université de Montréal, Montréal QC Canada \and 
 University of Washington, Tacoma WA USA \and
 Ghent University, Ghent Belgium \and
HEC Montréal, Montréal QC Canada
\\
\email{\{sunjiaao,farnadig\}@mila.quebec} \\
\email{\{sikha,mdecock\}@uw.edu}}
\maketitle              
\begin{abstract}
Users worldwide access massive amounts of curated data in the form of rankings on a daily basis. The societal impact of this ease of access has been studied and work has been done to propose and enforce various notions of fairness in rankings. Current computational methods for fair item ranking rely on disclosing user data to a centralized server, which gives rise to privacy concerns for the users. This work is the first to advance research at the conjunction of producer (item) fairness and consumer (user) privacy in rankings by exploring the incorporation of privacy-preserving techniques; specifically, differential privacy and secure multi-party computation. Our work extends the equity of amortized attention ranking mechanism to be privacy-preserving, and we evaluate its effects with respect to privacy, fairness, and ranking quality. Our results using real-world datasets show that we are able to effectively preserve the privacy of users and mitigate unfairness of items without making additional sacrifices to the quality of rankings in comparison to the ranking mechanism in the clear.
\end{abstract}

\section{Introduction}\label{SEC:intro}

Information systems, such as those used for search retrieval and recommendation, have become a ubiquitous part of our daily lives. These systems provide users with curated information such as content produced in social media or web pages, or organize the information in a way that is most relevant to users to aid their decision-making on what to buy, what to watch, or who to hire. These results are often outputted in the form of ranked lists. To maximize the utility of the system for each user, the ranked list is ordered by decreasing relevance (based on some score or rating) to the user. Users are then susceptible to paying most of their attention to the highest-ranked items in the ranked list, causing \textit{position bias} \cite{joachims2007search,joachims2007evaluating}. For a single ranked list (generated for a single user or a single query), the attention given to each item would decrease at a faster rate than relevance, \textit{i.e.}, lower-ranked items become disadvantaged by receiving disproportionately less attention relative to their relevance. When a sequence of ranked lists is generated in this manner, higher-ranked items reap increasingly disproportionately more attention relative to their relevance. This results in economic and social impacts on major stakeholders--the item providers, \textit{a.k.a.} producers--of the information systems, leading to economic disparity, bias toward underrepresented producers, and unhealthy markets \cite{kay2015unequal,liu2019real}.

Methods for mitigating unfairness in rankings have been proposed in the literature \cite{yang2017measuring,mehrotra2018towards,singh2019policy,zehlike2017fa,morik2020controlling}. In our work, we focus on post-processing techniques that reorder ranked lists to distribute exposure fairly \cite{singh2018fairness,biega2018equity}. 
These methods assume that the rankings or preferences of the users are available to a central entity that can then apply post-processing techniques to achieve fairness for the items. Centralizing user data can lead to privacy leakage or even intentional privacy violations, such as when the data is routinely sold when companies undergo bankruptcy \cite{canny2002collaborative}. The growing awareness of the need for more stringent user privacy protections, as well as the requirement to comply with regulations such as the GDPR\footnote{European General Data Protection Regulation \url{https://gdpr-info.eu/}}, is prompting the increased use of privacy-enhancing technologies (PETs) in recommender systems \cite{ge2021privitem2vec,yang2018privacy,ammad2019federated,chai2020secure,wu2021fedgnn}.
To the best of our current knowledge, no works address the challenge of achieving fairness for producers (\textit{a.k.a.} items) while preserving the privacy of consumers (\textit{a.k.a.} users) in ranking systems. In this paper, we explore this under-researched area in the literature.

We address the problem where each user (client) $u_l$ has a list $\rho_l$ of items $d_{1}^l, d_{2}^l, \ldots, d_{i}^l, \ldots, d_{n}^l$ that was generated in a privacy-preserving manner and ranked according to relevance. During a post-processing phase for bias mitigation, we wish to alter every user's ranking $\rho_l$ to optimize for equity of amortized attention~\cite{biega2018equity} without requiring any user disclosure of $\rho_l$ to a centralized entity.
One could achieve this goal with secure multi-party computation (MPC), \textit{i.e.}, by having each client encrypt their data to send to a set of MPC servers, having the MPC servers perform all computations needed for reranking over the encrypted data, and then return the results to the client to decrypt. While this approach would preserve end-to-end privacy and produce the same rankings as the bias mitigation method from Biega et al.~\cite{biega2018equity} yields in the clear (\textit{i.e.}, without any encryption), the computational and communication overhead would be prohibitively large for practical applications. We therefore propose a much more scalable approach by combining MPC to preserve input privacy with differential privacy (DP) to preserve output privacy. In our solution, MPC servers store intermediate results and then perturb them with DP noise, which clients subsequently use to perform local computations. We demonstrate the applicability of our proposed solution by experimenting on three real-world datasets. We empirically show that fairness for items can be improved while ensuring the privacy of users, without making additional sacrifices to the ranking quality.

\section{Preliminaries}\label{SEC:prelims}
\subsection{Fairness in Ranking}\label{SEC:prelimsFair}
To maximize utility, ranking algorithms generate lists of items sorted by their predicted level of relevance to a query or user. The most relevant items are positioned at the top of the list and receive the most exposure. Fairness in rankings is concerned with distributing exposure to the ranked items in order to mitigate the consequences of position bias. Many fairness notions have been proposed in this context \cite{yang2017measuring,zehlike2017fa,celis2017ranking,sapiezynski2019quantifying,singh2018fairness}. In this paper, we consider the fairness notion based on the attention received by items, which is dependent on the items' exposure, as proposed by Biega et al.~\cite{biega2018equity}. 
\subsubsection{Equity of Amortized Attention (EOAA) }
Biega et al. introduced the fairness notion of \textit{equity of amortized attention} to achieve \textit{amortized individual fairness} for a set of items $\mathcal{D} =\{d_{1}, d_{2}, \ldots, d_{i}, \ldots, d_{n}\}$ appearing in a sequence of relevance-based rankings $\rho_1,\rho_2, \ldots,\rho_l,\ldots,\rho_L$ for users  $u_1, u_2, \ldots, u_l, \ldots, u_L$, respectively \cite{biega2018equity}. The position of item $d_{i}$ in ranking $\rho_l$ influences the amount of attention that $d_{i}$ receives. EOAA is achieved if each item $d_{i}$ receives cumulative attention $A_i$ proportional to its cumulative relevance $R_i$, when amortized over a sequence of rankings. Biega et al. define a measure for this notion of fairness by taking the sum of the absolute differences between $A_i$ and $R_i$ for $i=1\dots n$ as shown in Eq.~\ref{eq:u_metric}:
 \begin{equation}\label{eq:u_metric}
     \textit{unfairness}(\rho_1, ..., \rho_L) = \sum_{i=1}^{n} \left| A_i - R_i \right| 
     = \sum_{i=1}^{n} \left| \sum_{l=1}^{L} a_i^l - \sum_{l=1}^{L} r_i^l \right|.
 \end{equation}
For each item $d_i$, $r_i^l$ is its relevance score for user $u_l$, and $a_i^l$ is the amount of attention it receives in ranking $\rho_l$. 
Biega et al. proposed to improve the EOAA of a sequence of relevance-based rankings $\rho_1,\rho_2, \ldots,\rho_l,\ldots,\rho_L$ by reranking each of the rankings sequentially to produce $\rho^*_1,\rho^*_2, \ldots,\rho^*_l,\ldots,\rho^*_L$. To rerank $\rho_l$, taking into account the previously computed rerankings $\rho^*_1,\rho^*_2, \ldots,\rho^*_{l-1}$, the following post-processing linear program (ILP) is solved:
\begin{eqnarray}\label{eq:ilp}
        \text{Minimize } \sum_{i=1}^{n} \sum_{j=1}^{n} \left| A_i^{l-1} + \hat{w}_j - (R_i^{l-1} + \hat{r}_i^l) \right| \cdot X_{i, j} \label{EQ:objective}\\    
        \text{Subject to } \sum_{j=1}^{k} \sum_{i=1}^{n} \frac{2^{\hat{r}_i^l}-1}{\log_2(j+1)} X_{i, j} \geq \theta \cdot DCG(\rho_l)@k \label{EQ:constraint} \\
        X_{i, j} \in \{0, 1\}, \forall_{i,j} \mbox{\ and\ }  \sum_i X_{i, j} = 1, \forall_j \mbox{\ and\ } \sum_j X_{i, j} = 1, \forall_i. \label{EQ:moreconstraints}
\end{eqnarray}
\noindent
This ILP solves for $n^2$ decision variables $X_{i, j}$ that represent the reranking $\rho_l^*$ of the items in ranking $\rho_l$.
$A_i^{l-1}$ and $R_i^{l-1}$ are the accumulated attention and relevance of item $d_i$, respectively, over rerankings $\rho^*_1,\rho^*_2, \ldots,\rho^*_{l-1}$. $\hat{w}_j$ is the normalized attention weight of placing an item at position $j$, calculated as $\hat{w}_j = w_j / \sum_{t=1}^n w_t$, based on the geometric attention model $w_j = 0.5(0.5)^{j-1}$ assumed in \cite{biega2018equity}.
$\hat{r}_i^l$ is the normalized relevance score of $r_i^l$, calculated as $\hat{r}_i^l = \left. \frac{r_i^l - r_{\min}}{r_{\max} - r_{\min}} \middle/ \sum_{t=1}^n \frac{r_t^l - r_{\min}}{r_{\max} - r_{\min}} \right.$. $r_{\max}$ and $r_{\min}$ are the maximum and minimum relevance scores, respectively, that a user can have for an item.
The coefficient $\left| A_i^{l-1} + w_j - (R_i^{l-1} + r_i^l) \right|$ is the unfairness measure of placing an item $d_i$ at position $j$ in the current $l^{th}$ reranking. The constraint in Eq.~\ref{EQ:constraint}
bounds the quality of the first $k$ items in the reranking in terms of its discounted cumulative gain (DCG), such that it is no lower than $0 \leq \theta \leq 1$ times the DCG of the top-$k$ items of the original relevance-based ranking. The constraints in Eq.~\ref{EQ:moreconstraints} specify that the $n^2$ decision variables $X_{i, j}$ are binary, and that there is a single 1 per $j$ for all $i$'s, and a single 1 per $i$ for all $j$'s. $X_{i, j} = 1$ indicates item $d_i$ is placed in position $j$, and $X_{i, j} = 0$ otherwise.

The quality of reranking $\rho^*_l$ is measured in terms of its divergence from the original relevance-based ranking $\rho_l$. This is quantified as $\textit{NDCG}(\rho_l, \rho_l^*)$ in Eq.~\ref{eq:ndcg_metric}, where $DCG(\rho^*_l)$ is normalized by $DCG(\rho_l)$.
\begin{equation}\label{eq:ndcg_metric}
    \textit{NDCG}(\rho_l,\rho^*_l) = \frac{DCG(\rho^*_l)}{DCG(\rho_l)}
\end{equation}
The maximum NDCG value is 1, and occurs when either $\rho_l = \rho^*_l$, or if items of equal relevance scores are shuffled with each other.

\subsection{Privacy-Enhancing Technologies (PETs)}\label{SEC:prelimsPrivacy}
\subsubsection{Differential Privacy (DP) } 
DP provides formal guarantees that the result of computations on a dataset $D$ is negligibly affected by the participation of a single user, thereby offering privacy through plausible deniability \cite{dwork2014algorithmic}. 
Formally, a randomized algorithm $\mathcal{F}$ provides $\epsilon$-DP if for all pairs of neighboring datasets $D$ and $D'$ (i.e.~datasets that differ in one entity), and for all subsets $S$ of $\mathcal{F}$'s range
\begin{equation}\label{DEF:DP}
\mbox{P}(\mathcal{F}(D) \in S) \leq e^{\epsilon} \cdot \mbox{P}(\mathcal{F}(D') \in S),
\end{equation}
where $\epsilon$ is the privacy budget or privacy loss. The smaller the value of $\epsilon$, the stronger the privacy guarantees. An $\epsilon$-DP algorithm $\mathcal{F}$ is usually created out of an algorithm $\mathcal{F^*}$ by adding noise that is proportional to the \textit{sensitivity} of $\mathcal{F^*}$, in which the sensitivity measures the maximum impact a change in the underlying dataset can have on the output of $\mathcal{F^*}$. In our paper, $\mathcal{F^*}$ performs aggregation of the relevance scores and the attention weights of items across many users. The traditional DP paradigm--global DP--assumes a central curator who collects data from the users and injects controlled noise either to the inputs, outputs, or both when revealing the computed aggregation. Although this model provides \textit{output privacy}, it requires users to disclose their private data with a central entity. To remove this need for sensitive user information disclosure, in this paper we emulate the central curator from the global DP paradigm with a set of untrusted servers running MPC protocols (see below).

\subsubsection{Secure Multi-Party Computation (MPC) }
MPC \cite{damgard} is an umbrella term for cryptographic approaches that enable
computations over encrypted data. We follow the MPC as a service paradigm in which each data holder encrypts each value $x$ from its input data by converting it into so-called secret shares and subsequently distributes these shares among a set of MPC servers. While the original value $x$ can be trivially reconstructed when all shares are combined, none of the MPC servers by themselves learns anything about the value of $x$. Next, the MPC servers jointly execute protocols to perform computations over the secret shared values to obtain a secret sharing of the desired output value (in our paper, an aggregated unfairness measure perturbed with noise to provide DP). MPC is concerned with the protocol execution coming under attack by an adversary, which may corrupt one or more of the parties to learn private information or cause the result of the computation to be incorrect. The MPC protocols that we use in this paper are designed to prevent such attacks from being successful, and they are mathematically proven to guarantee \textit{input privacy} and correctness.

\textbf{Threat Model } An adversary can corrupt any number of parties. In a \textit{dishonest-majority} setting, half or more of the parties may be corrupt, while in an \textit{honest-majority} setting, more than half of the parties are honest (not corrupted). Furthermore, the adversary can be a \textit{semi-honest} or a \textit{malicious} adversary. While a party corrupted by a semi-honest or `passive' adversary follows the protocol instructions correctly but tries to obtain additional information, parties corrupted by malicious or `active' adversaries can deviate from the protocol instructions. The protocols we use in this paper are sufficiently generic to be used in dishonest-majority as well as honest-majority settings, with
passive or active adversaries. This is achieved by changing
the underlying MPC scheme to align with the desired
security setting.
 
MPC computations are commonly done on integers modulo $q$, \textit{i.e.},~in the ring $\mathbb{Z}_{q}$. For instance, in a well-known dishonest majority 2-party (2PC) computation setting with passive adversaries, a data holder converts its input data into $x = x_0 + x_1 \mod q$ and sends $x_0$ to MPC server $S_0$, and $x_1$ to $S_1$. We use $[\![x]\!] = (x_0,x_1)$ as a shorthand for a secret sharing of $x$ throughout the paper. If the servers have received secret shares $y_0$ and $y_1$ of a value $y$ from another data holder, then the servers can compute a secret sharing of $x+y$ as $[\![x+y]\!] = (x_0+y_0,x_1+y_1)$ without learning the values of $x$, $y$, or their sum. Besides addition, we use an MPC protocol $\pLAP$ that enables the MPC servers to sample secret sharings of noise drawn from a Laplace distribution. $\pLAP$ is substantially more complex and involves communication between the MPC servers in addition to operations on their own local shares. See \cite{pentyala2022privfairfl} for details.

\section{Related Work}\label{SEC:related}

\subsection{Fairness in Rankings }
Most work in fairness in rankings has been studied in the context of fairly distributing exposure to the elements of the ranked list. The elements of the ranked list could be people or items (\textit{e.g.}, content, products, places). Much of the work involving exposure has centered around group fairness, where exposure should ideally be distributed equally among different groups defined by their protected attributes (such as gender or race). For example, Yang and Stoyanovich~\cite{yang2017measuring} proposed extending the traditional fairness concept of statistical parity to the ranking system, where being a member of a protected group does not influence a person’s position in a ranking. Zehlike and Castillo~\cite{zehlike2017fa} proposed a fair top-$k$ ranking algorithm following the fairness notion of affirmative action, where a minimum number of protected group members are guaranteed in every top-$k$ ranking (top 10, top 20, etc.). Celis et al.~\cite{celis2017ranking} proposed an algorithm addressing the constrained ranking maximization problem, where there is a limit to the number of sensitive items per protected group in the ranking, and no one group dominates. Sapiezynski et al.~\cite{sapiezynski2019quantifying} also used statistical parity, but used a geometric distribution to model user attention as an analogue to exposure. Singh and Joachims~\cite{singh2018fairness} introduced fairness of exposure in rankings and proposed to use linear programming to optimize for the maximum utility in a ranking, subject to group fairness constraints. However, limited studies have been done on individual item fairness in rankings. One such work is Biega et al.’s~\cite{biega2018equity} proposal of equity of attention, which aims to achieve amortized fairness over a sequence of rankings by distributing attention proportional to item relevance. Our work extends Biega et al.’s~\cite{biega2018equity} reranking approach by implementing privacy-preserving measures at various stages of the ranking mechanism.

\subsection{Privacy-Preserving Ranking Systems }
Various PETs have been used in previous works on privacy-preserving learning-to-rank (LTR) systems. Dehghani et al.~\cite{dehghani2017share} used mimic learning, where only a model trained on the sensitive data is shared, and not the data itself. Furthermore, Laplace noise is used during aggregation as part of the DP guarantee. Yang et al.~\cite{yang2018privacy} used the information-theoretic privacy approach, which involves obfuscating data in accordance with a data distortion budget. Kharitonov~\cite{kharitonov2019federated} used a federated learning setup with evolution strategies optimization, and then incorporated local DP. Wang et al.~\cite{wang2021federated} extended Kharitonov’s work to larger datasets and found a substantial loss in utility compared to other non-private online learning-to-rank systems. Wang et al.~\cite{wang2021effective} then used a federated learning setup and local DP, similar to Kharitonov~\cite{kharitonov2019federated}, but incorporated a pairwise differentiable gradient descent (PDGD) optimization approach instead. Ge et al.~\cite{ge2021privitem2vec} incorporated the Paillier homomorphic encryption algorithm in their PrivItem2Vec model. Among these previous works in privacy-preserving ranking systems, none have exploited the use of MPC together with DP.

\subsection{Fair and Privacy-Preserving Ranking Systems}
It is evident that both fairness in rankings and privacy-preserving methods in ranking systems have been well-studied; however, there is a dearth of research at the intersection of fairness and privacy in ranking systems. Resheff et al.~\cite{resheff2018privacy} used privacy-adversarial training in their recommender system to obtain user representations that obfuscate users’ sensitive attributes. 
This approach focuses on improving group fairness and preserves some user privacy by way of preventing implicit private information attacks. Sato~\cite{sato2022private} proposed a local ranking system framework that is independent from the centralized recommender system, where users can post-process the rankings they receive by themselves by setting their own fairness constraints to their preferences. Their privacy-preserving method is in each user developing their own recommender system. However, this is very computationally expensive and many users' devices do not have the processing power to maintain these systems. Both of these works address privacy and fairness with respect to the users’ protected attributes, yet to the best of our knowledge, there has been no work so far that addresses privacy and fairness with respect to the amount of attention items receive in relation to their relevance.

\section{Methodology} 
\label{SEC:method}

\subsection{Problem Description}
We consider a regression-style recommendation model $\mathcal{M}$ that is trained in a privacy-preserving manner (such as \cite{wu2021fedgnn,kharitonov2019federated}), \textit{i.e.}, the model does not leak any information about the training data. $\mathcal{M}$ is deployed at each user $u_l$ to predict their relevance scores $\vectorbold{r}^l = (r_1^l, r_2^l, \dots, r_i^l, \dots, r_n^l)$ on a global set of items $\mathcal{D}$, where $r_i^l = \mathcal{M}(u_l, d_i)$. In this process, the users' raw data--such as their preferences, demographics, embeddings, etc.--are not disclosed to anyone. User $u_l$'s relevance-based ranking $\rho_l$ is its list of items sorted in decreasing order of their relevance scores.

The post-processing technique described in Section \ref{SEC:prelimsFair} assumes that a central server $S$ accesses the relevance-based rankings $\rho_1, \rho_2, \ldots, \rho_l, \ldots, \rho_L$ of all users and reranks them into $\rho_1^*,\rho_2^*, \ldots, \rho_l^*, \ldots, \rho_L^*$ to achieve \textit{equity of amortized attention} without losing the ranking utility beyond the set threshold. 
This setup, which we refer to as the \textit{centralized setup}, causes leakage of sensitive user information to the central server $S$, including:
\begin{enumerate}[noitemsep,topsep=0pt,leftmargin=*,label=(\subscript{P}{{\arabic*}})]
    \item the preference of the user for all items in the form of relevance scores, 
    \item the top-$k$ items that the user is most likely to be interested in, and
    \item the order of the top-$k$ items the user is most likely to be interested in.    
\end{enumerate}

Below we describe how we adapt the above post-processing method to address privacy issues ($P_1$)--($P_3$) in order to achieve individual fairness for the items and preserve the privacy of the users.

\begin{figure}
    \centering
    \includegraphics[width=1\linewidth]{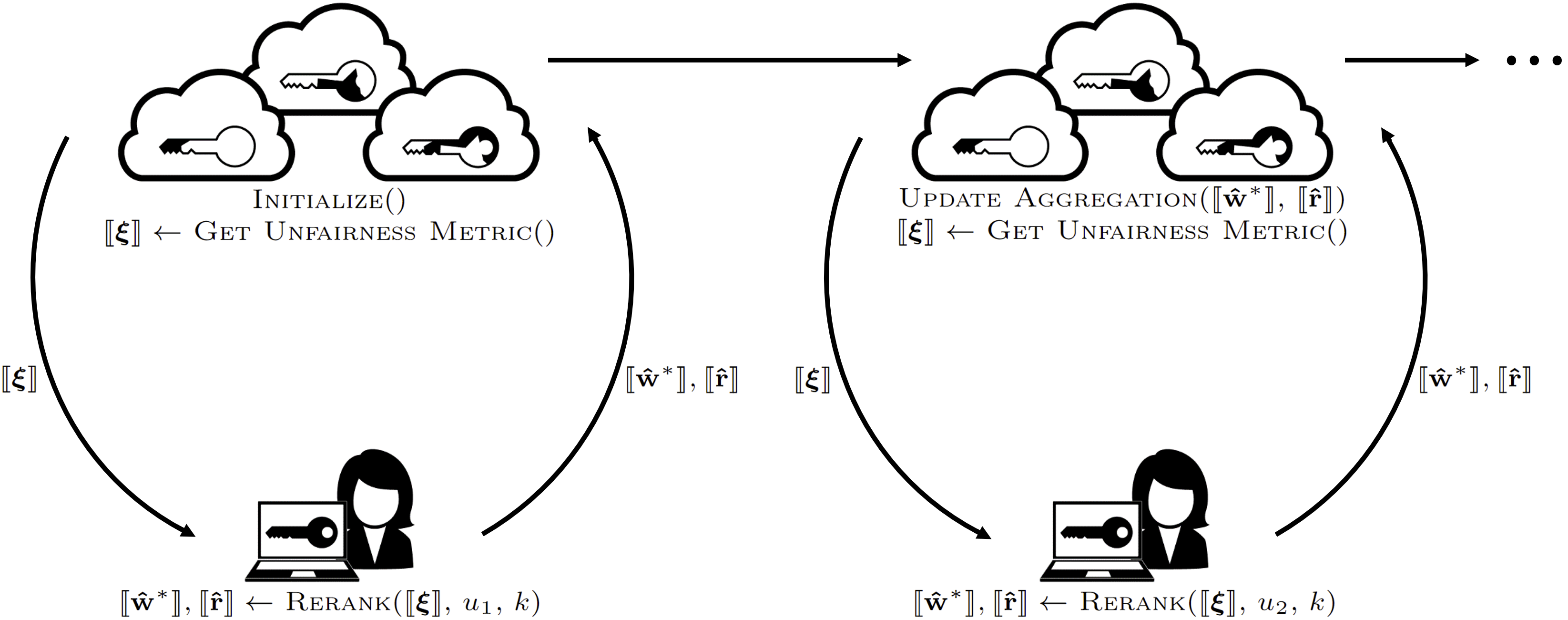}
    \caption{Flow diagram of privacy-preserving fair item ranking algorithm\protect\footnotemark}
    \label{fig:our_approach}
\end{figure}
\footnotetext{Figure adapted from \url{https://sepior.com/mpc-blog}}

\subsection{Proposed Method}
The key observation underlying our method is that each user $u_l$ has all the information needed to solve the ILP to rerank their original ranking $\rho_l$ into $\rho^*_l$, except for the values of $A_i^{l-1}$ and $R_i^{l-1}$ in Eq.~\ref{EQ:objective}. 
$A_i^{l-1}$ and $R_i^{l-1}$ are the accumulated attention and relevance of item $d_i$, respectively, over rerankings $\rho^*_1,\rho^*_2, \ldots,\rho^*_{l-1}$, \textit{i.e.},~the values of $A_i^{l-1}$ and $R_i^{l-1}$ depend on sensitive information from users $u_1, \ldots, u_{l-1}$ which we neither want to disclose to user $u_l$ nor to a central server (as is the case in the centralized setup). In our solution, we therefore maintain encrypted versions of $A_i^{l-1}$ and $R_i^{l-1}$ which initially are 0 (see Procedure \textsc{Initialize} in Algorithm \ref{alg:mpc}), and are updated in a privacy-preserving manner each time a user completes their ILP computation. More precisely, no entity knows by itself at any point what the current values of $A_i^{l-1}$ and $R_i^{l-1}$ are. Instead, these values are split into secret shares and distributed over MPC servers who can jointly perform operations to update the shares, without ever learning the true values of the inputs or the results of the computations. Figure \ref{fig:our_approach} illustrates the high-level flow of our solution, and Algorithm \ref{alg:mpc} presents the pseudo-code, which we explain in more detail below. We use $\vectorbold{A}^{l-1}$ and $\vectorbold{R}^{l-1}$ to denote the vectors $[A_1^{l-1}, \ldots, A_n^{l-1}]$ and $[R_1^{l-1}, \ldots, R_n^{l-1}]$, respectively.

The users update their rankings in sequence $l = 1 \ldots L$. When the sequence reaches user $u_l$, $u_l$ first requests the current vector of differences $\vectorbold{A}^{l-1} - \vectorbold{R}^{l-1}$ from the MPC servers, prompting the MPC servers to compute a secret sharing $[\![\vectorbold{A}^{l-1} - \vectorbold{R}^{l-1}]\!]$ from their local secret shares of $\vectorbold{A}^{l-1}$ and $\vectorbold{R}^{l-1}$  (Procedure \textsc{Get Unfairness Metric}). In principle, each MPC server could send their secret shares of $\vectorbold{A}^{l-1} - \vectorbold{R}^{l-1}$ to user $u_l$, which $u_l$ could combine to construct the value of $\vectorbold{A}^{l-1} - \vectorbold{R}^{l-1}$. However, although $\vectorbold{A}^{l-1} - \vectorbold{R}^{l-1}$ consists of aggregated information only, this value could still leak information about the previous users $u_1,\ldots,u_{l-1}$ to user $u_l$, especially if $u_l$ is one of the first to rerank. In our solution, this privacy loss is mitigated by having the MPC servers perturb the value at each index of $[\![\vectorbold{A}^{l-1} - \vectorbold{R}^{l-1}]\!]$ with Laplace noise \textit{before} sending it to user $u_l$ (Procedure \textsc{Get Unfairness Metric}). We denote this perturbed secret sharing as $[\![\boldsymbol{\xi}]\!]$. The DP guarantee that the MPC servers provide in this manner is that the probability of returning any specific value of $\boldsymbol{\xi}$ is very similar to the probability of returning that value if the data of a previous user $u_i$ ($i=1\ldots l-1$) would have been left out of the computation of  $[\![\vectorbold{A}^{l-1} - \vectorbold{R}^{l-1}]\!]$ (see Eq.~\ref{DEF:DP}). This entails that the value of $\boldsymbol{\xi}$ returned to user $u_l$ does not leak information about the users who computed their rerankings before it was $u_l$'s turn. To this end, the MPC servers generate secret shares of Laplace noise using the MPC-protocol $\pLAP$ for secure sampling from a Laplace distribution as described by Pentyala et al.~\cite{pentyala2022privfairfl} and add these secret shares to $[\![\vectorbold{A}^{l-1} - \vectorbold{R}^{l-1}]\!]$, effectively emulating the global DP paradigm without having to rely on a central trusted aggregator. We provide more information about the scale parameter $b$ for the Laplace noise below.

\begin{algorithm}
\scriptsize{
\caption{Privacy-Preserving Fair Item Ranking}
\label{alg:mpc} 
\begin{multicols}{2}
\begin{algorithmic}
\State \textbf{\underline{Achieving EOAA privately over $L$ users}}
\State $n \leftarrow$ number of items per ranking
\State $L \leftarrow$ number of rankings to rerank
\State $k \leftarrow$ number of items in quality constraint Eq.~\ref{EQ:constraint}
\State $\vectorbold{w} \leftarrow$ attention weights
\State $\vectorbold{\hat{w}} \leftarrow \Call{Normalize}{\vectorbold{w}}$
\State \Call{Initialize}{\hphantom{}}
\For{\textbf{each} $u_{l}$, \textbf{where} $l=1 \ldots L$}
\State $[\![\boldsymbol{\xi}]\!] \leftarrow$ \Call{Get Unfairness Metric}{\hphantom{}}
\State $[\![\vectorbold{\hat{w}}^*]\!], [\![\vectorbold{\hat{r}}]\!] \leftarrow$ \Call{Rerank}{$[\![\boldsymbol{\xi}]\!]$, $u_{l}$, $k$}
\State \Call{Update Aggregation}{$[\![\vectorbold{\hat{w}}^*]\!]$, $[\![\vectorbold{\hat{r}}]\!]$}
\EndFor

\Statex

\State \textbf{\underline{User $u_{l}$ subroutine}}
\Procedure {Rerank}{$[\![\boldsymbol{\xi}]\!], u_{l}, k$}
\State $\boldsymbol{\xi} \leftarrow  (\epsilon / L) \sum [\![\boldsymbol{\xi}]\!]$
\State $\vectorbold{r} \leftarrow \mathcal{M}(u_l,\mathcal{D})$
\State $\vectorbold{\hat{r}} \leftarrow \Call{Normalize}{\vectorbold{r}}$
\State $\vectorbold{s} \leftarrow$ \Call{ILP}{$\boldsymbol{\xi}, \vectorbold{\hat{r}}, k$}
\State $\vectorbold{\hat{w}}^* \leftarrow \vectorbold{\hat{w}}[\vectorbold{s}]$
\State Return $[\![\vectorbold{\hat{w}}^*]\!], [\![\vectorbold{\hat{r}}]\!]$
\EndProcedure

\columnbreak

\State \textbf{\underline{MPC servers' subroutines}}
\Procedure {Initialize}{\hphantom{}}
\State $\mathrm{\Delta} f \leftarrow$ sensitivity calculated from Eq.~\ref{EQ:sensitivity}
\State $\epsilon \leftarrow$ privacy budget
\State $[\![\vectorbold{A}]\!] \leftarrow [\![0]\!]$
\State $[\![\vectorbold{R}]\!] \leftarrow [\![0]\!]$
\EndProcedure
\Statex

\Procedure {Get Unfairness Metric}{\hphantom{}}
\State //MPC protocol for global DP
\State $b \leftarrow \mathrm{\Delta} f / (\epsilon / (n \cdot L))$
\State $[\![\boldsymbol{\xi}]\!] \leftarrow [\![\vectorbold{A} - \vectorbold{R}]\!] + \pLAP(b)$
\State Return $[\![\boldsymbol{\xi}]\!]$
\EndProcedure
\Statex

\Procedure {Update Aggregation}{$[\![\vectorbold{\hat{w}}^*]\!], [\![\vectorbold{\hat{r}}]\!]$}
\State //MPC protocol to perform aggregation
\State $[\![\vectorbold{A}]\!] \leftarrow [\![\vectorbold{A}]\!] + [\![\vectorbold{\hat{w}}^*]\!]$ 
\State $[\![\vectorbold{R}]\!] \leftarrow [\![\vectorbold{R}]\!] + [\![\vectorbold{\hat{r}}]\!]$ 
\EndProcedure
\Statex

\Statex

\end{algorithmic}
\end{multicols}
}
\end{algorithm}

Once user $u_l$ has received secret shares of $\boldsymbol{\xi}$ from each MPC server, $u_l$ can sum up the secret shares to get $\vectorbold{A}^{l-1} - \vectorbold{R}^{l-1} + \pLAP(b)$, using this as a proxy for $\vectorbold{A}^{l-1} - \vectorbold{R}^{l-1}$, and then proceed to solve the ILP program in Sec.~\ref{SEC:prelimsFair} (Procedure \textsc{Rerank}) for $X_{i, j}$ ($i=1\dots n$ and $j=1\dots n$). In our implementation, we scaled $\vectorbold{A}^{l-1} - \vectorbold{R}^{l-1} + \pLAP(b)$ by a positive factor $\epsilon/L$ so that the solution of the ILP is easier to compute. We note that scaling by a positive factor does not affect the outcome in $X_{i, j}$. $X_{i, j}$ can then translate to a vector $\vectorbold{s}$ of indices to reorder the normalized attention weights $\vectorbold{\hat{w}}$. $\vectorbold{\hat{w}}^*$ is the vector of attention weights distributed to each item at each index in the order of $\vectorbold{s}$. Thus, $\vectorbold{\hat{w}}^*$ and $\vectorbold{\hat{r}}$ make up the reranking $\rho^*_l$. User $u_l$ subsequently encrypts the values in these vectors by splitting them into secret shares $[\![\vectorbold{\hat{w}}^*]\!]$ and $[\![\vectorbold{\hat{r}}]\!]$, and distributes the shares among the MPC servers. This enables the MPC servers to update their secret shares  of the aggregated values to $[\![\vectorbold{A}^{l}]\!]$ and $[\![\vectorbold{R}^{l}]\!]$ (Procedure \textsc{Update Aggregations}), which they will need when the next user, $u_{l+1}$, makes a request. The whole process is repeated until all users have completed their reranking.

\textbf{Scale parameter $b$}.
To make our algorithm $\epsilon$-DP, the MPC servers answer each query by adding noise to the true aggregate. To this end, the MPC servers sample noise from a Laplace distribution with mean 0 and scale $b = \mathrm{\Delta} f/ \epsilon'$ in which $\mathrm{\Delta} f$ denotes the sensitivity and $\epsilon'$ the privacy budget per query. Appropriate values for these parameters are described next.
In Algorithm \ref{alg:mpc}, each user $u_l$ ($l=1\ldots L$) queries the MPC servers for aggregated information $A_i^{l-1} - R_i^{l-1}$ about each item $d_i$ ($i= 1\ldots n$) through the Procedure \textsc{Get Unfairness Metric}. The total number of queries to be answered by the MPC servers is in other words $n \cdot L$. These queries are executed against overlapping datasets, as $u_l$ queries the aggregated information of users $u_1, \ldots, u_{l-1}$; $u_{l+1}$ queries the aggregated information of users $u_1, \ldots, u_{l}$, etc. Given a total privacy budget $\epsilon$, we therefore allocate $\epsilon'=\epsilon/(n \cdot L)$ per query.

$\mathrm{\Delta} f$ is the sensitivity computed for the aggregate ${A_i}^{l-1} - {R_i}^{l-1}$, and is given by Eq.~\ref{EQ:sensitivity} that computes the maximum value that can be contributed to this aggregate by any single user.
\begin{align}
    \hat{r}_{\min} &= \left. \frac{r_{\min} - r_{\min}}{r_{\max} - r_{\min}} \middle/ \left( \frac{r_{\min} - r_{\min}}{r_{\max} - r_{\min}} + (n - 1) \left( \frac{r_{\max} - r_{\min}}{r_{\max} - r_{\min}} \right) \right) \right. = 0 \label{EQ:changeinrmin}\\
    \hat{r}_{\max} &= \left. \frac{r_{\max} - r_{\min}}{r_{\max} - r_{\min}} \middle/ \left( \frac{r_{\max} - r_{\min}}{r_{\max} - r_{\min}} + (n - 1) \left( \frac{r_{\min} - r_{\min}}{r_{\max} - r_{\min}} \right) \right) \right. = 1 \label{EQ:changeinrmax}\\
    \hat{w}_{\min} &= \frac{w_{n}}{\sum_{j = 1}^n w_j} \label{EQ:changeinwmin}\\
    \hat{w}_{\max} &= \frac{w_{1}}{\sum_{j = 1}^n w_j} \label{EQ:changeinwmax}\\   
    \mathrm{\Delta} f &= \max(|\hat{w}_{\max} - \hat{r}_{\min}| , |\hat{w}_{\min} - \hat{r}_{\max}|) \label{EQ:sensitivity}
\end{align}

Eqs. \ref{EQ:changeinwmin}, \ref{EQ:changeinwmax} are based on the normalized geometric attention model and  Eqs. \ref{EQ:changeinrmin}, \ref{EQ:changeinrmax} are based on the range of the normalized relevance scores that the MPC servers may receive. Computation of $\mathrm{\Delta} f$ and $b$ is independent of the users' data and can be precomputed by one of the MPC servers in the clear, \textit{i.e.}, without encryption (see Procedure \textsc{Initialize}).

\section{Results}\label{SEC:results}

\subsection{Datasets}
We use three recommender system datasets in our experiments: Amazon Digital Music,\footnote{\url{http://jmcauley.ucsd.edu/data/amazon/links.html}} Book Crossing,\footnote{\url{http://www2.informatik.uni-freiburg.de/~cziegler/BX/}} and MovieLens-1M.\footnote{\url{https://grouplens.org/datasets/movielens/1m/}} Each dataset contains information about each user and item, and ratings that users gave to items. We detail the number of users, items, ratings, and the range of possible ratings of each dataset in Table~\ref{table:datasets}.

\begin{table}[h!]
    \caption{Statistics of datasets used to train each SVD model}
    \label{table:datasets}
    \centering
    \begin{tabular}{p{3.6cm}
        >{\raggedleft\arraybackslash}p{1.2cm}
        >{\raggedleft\arraybackslash}p{1.5cm}
        >{\raggedleft\arraybackslash}p{1.8cm}
        p{2.5cm}}
        \toprule
        Dataset & \# Users & \# Items & \# Ratings & ~~~~Rating Levels\\
        \midrule
        Amazon Digital Music & $478,235$ & $266,414$ & $836,006$ & ~~~~$1, 2, ..., 5$\\
        Book Crossing & $77,805$ & $185,973$ & $433,671$ & ~~~~$1, 2, ..., 10$\\ 
        MovieLens 1M & $6,040$ & $3,706$ & $1,000,209$ & ~~~~$1, 2, ..., 5$\\ 
        \bottomrule
    \end{tabular}
\end{table}

\subsection{Experimental Setup}
We trained a singular value decomposition (SVD) model\footnote{\url{https://surpriselib.com/}} for each dataset and predicted relevance scores for every user-item pair. We assume that these models were trained in a privacy-preserving manner.
For our experiments, we select $n=100$ items to rerank for all datasets, and rerank $L=3000$ users' rankings for the Amazon Digital Music and Book Crossing datasets, and all $L=6040$ users' rankings for the MovieLens-1M dataset. We use Gurobi\footnote{\url{https://www.gurobi.com/}} to solve the ILP in Eq.~\ref{EQ:objective}--\ref{EQ:moreconstraints}. We set $k=100$, the quality loss constraint to $\theta=0.8$, and calculate the sensitivity based on Eq.~\ref{EQ:sensitivity} to be $\mathrm{\Delta} f=1$ for all datasets. We perform an empirical analysis for privacy budget $\epsilon \in \{0.5, 1, 10, 100, 1000, 10000, 100000 \}$.

All MPC computations are done in the MP-SPDZ framework \cite{mpspdz} and performed over a ring $\mathbb{Z}_q$ with $q=2^{64}$. We perform experiments in a dishonest majority security setting with 2 computing parties (2PC) and passive adversaries.

We evaluate our approach in terms of unfairness (Eq.~\ref{eq:u_metric}) and utility (Eq.~\ref{eq:ndcg_metric})  and study the privacy-fairness-utility trade-offs.


\begin{figure}[t!] 
\begin{subfigure}{0.49\textwidth}
\includegraphics[width=\linewidth]{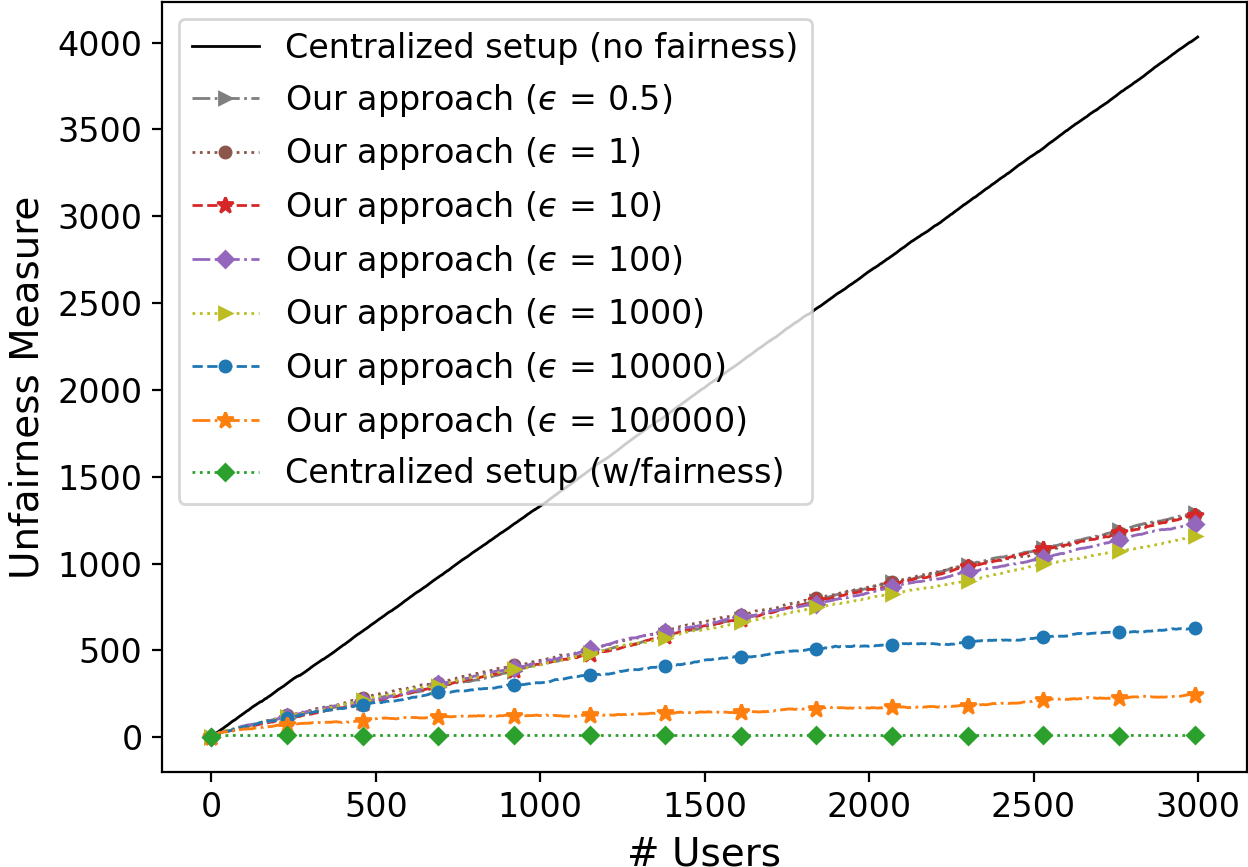}
\caption{Unfairness Measure on Amazon Digital Music} \label{fig:a}
\end{subfigure}\hspace*{\fill}
\begin{subfigure}{0.49\textwidth}
\includegraphics[width=\linewidth]{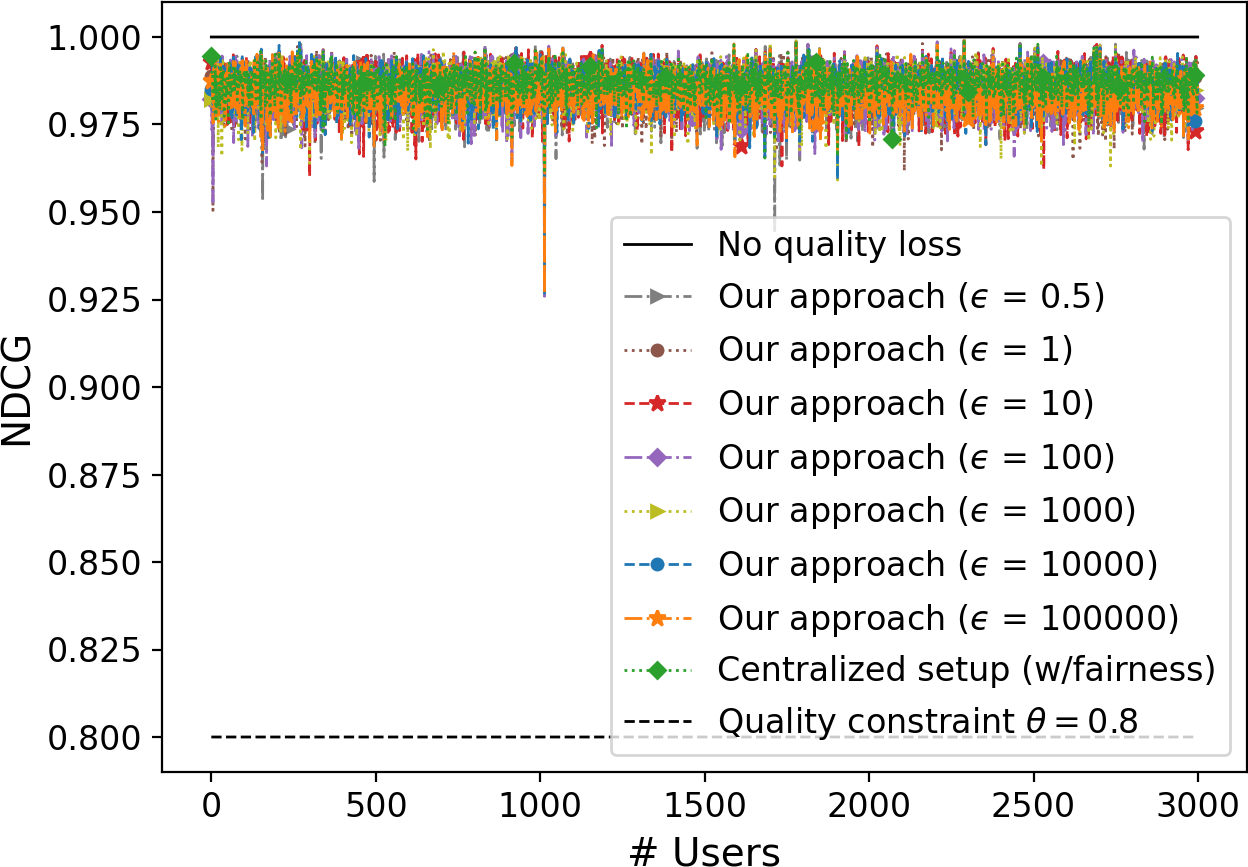}
\caption{Ranking Quality on Amazon Digital Music} \label{fig:b}
\end{subfigure}\hspace*{\fill}

\medskip
\begin{subfigure}{0.49\textwidth}
\includegraphics[width=\linewidth]{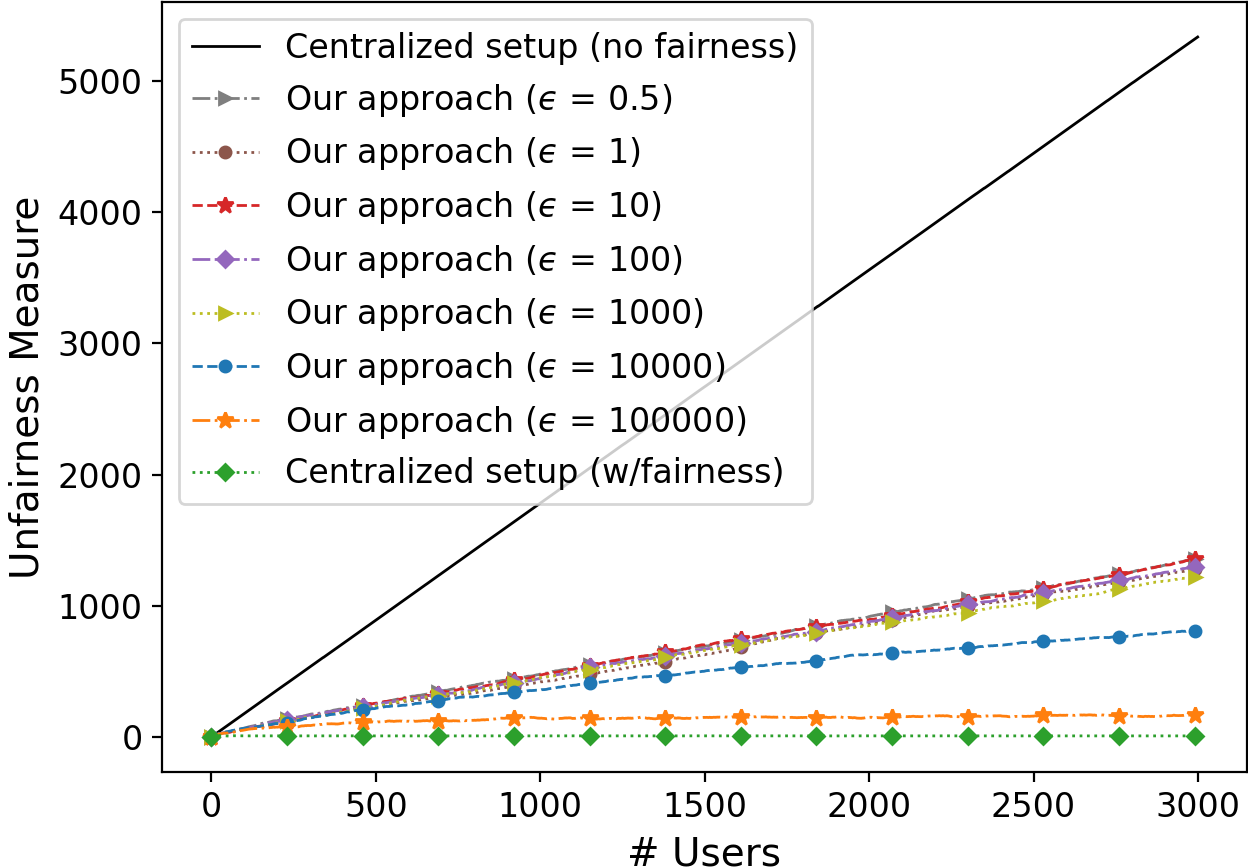}
\caption{Unfairness Measure on Book Crossing} \label{fig:c}
\end{subfigure}\hspace*{\fill}
\begin{subfigure}{0.49\textwidth}
\includegraphics[width=\linewidth]{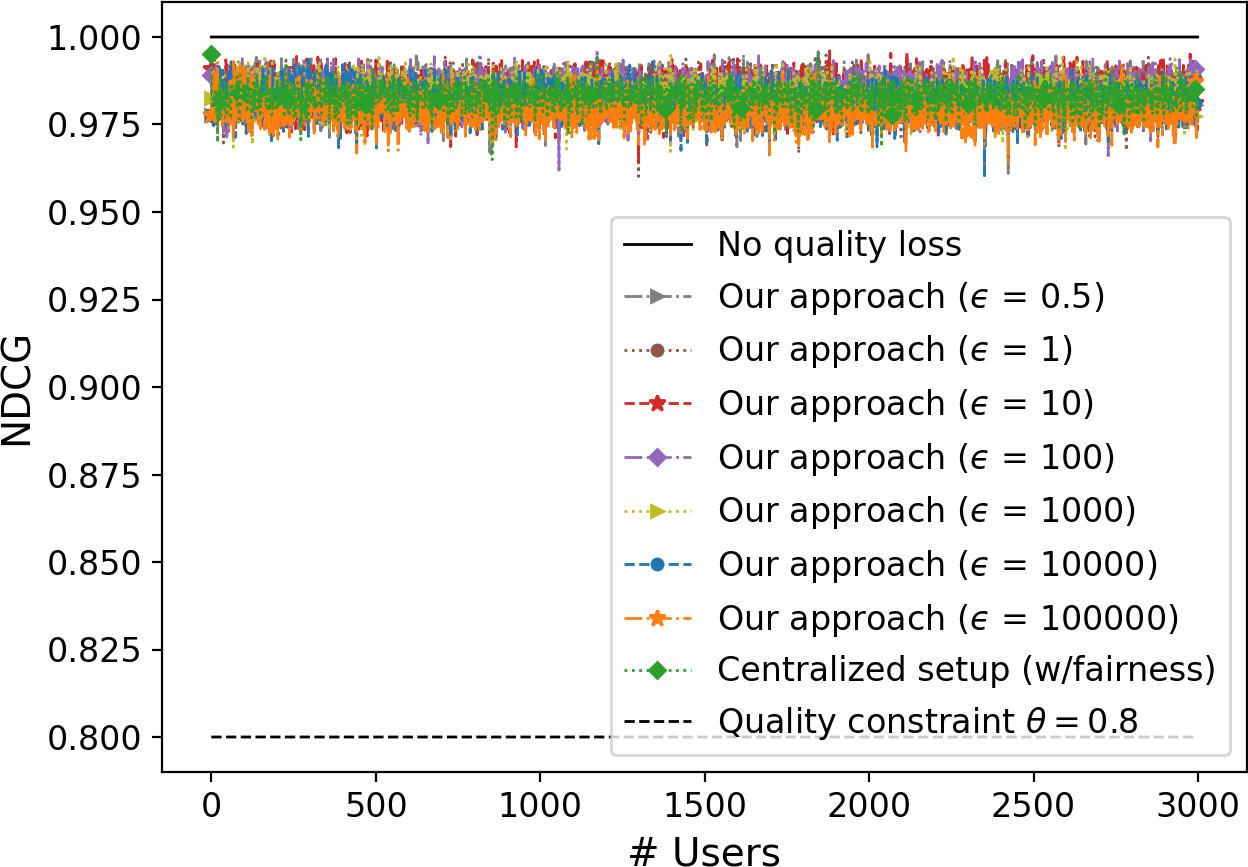}
\caption{Ranking Quality on Book Crossing} \label{fig:d}
\end{subfigure}\hspace*{\fill}

\medskip
\begin{subfigure}{0.49\textwidth}
\includegraphics[width=\linewidth]{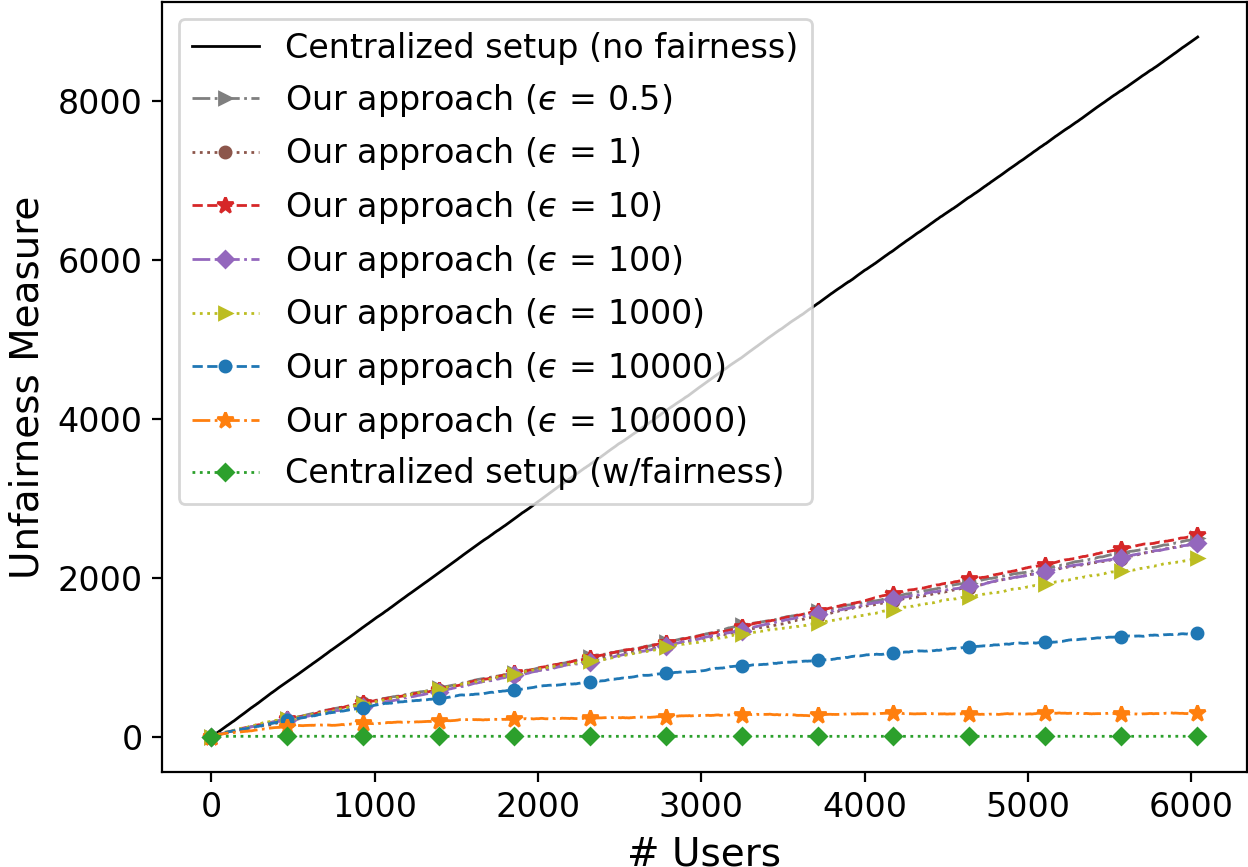}
\caption{Unfairness Measure on MovieLens-1M} \label{fig:e}
\end{subfigure}\hspace*{\fill}
\begin{subfigure}{0.49\textwidth}
\includegraphics[width=\linewidth]{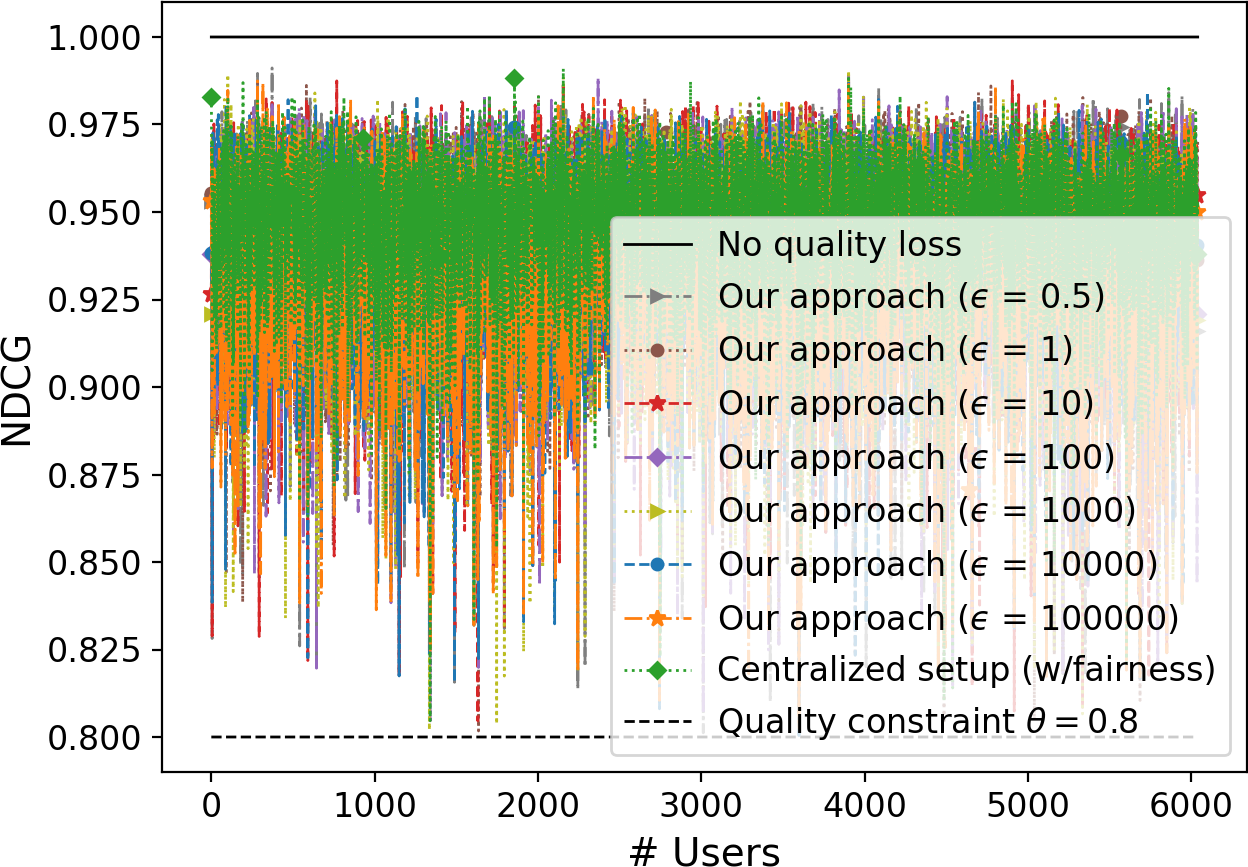}
\caption{Ranking Quality on MovieLens-1M} \label{fig:f}
\end{subfigure}\hspace*{\fill}

\caption{Model performance on each dataset} \label{fig:results}
\end{figure}

\subsection{Discussion}
\subsubsection{Fairness vs. Privacy Trade-offs }
We demonstrate the cost on item fairness when preserving the privacy of users in Figure \ref{fig:results} (left column; Figures \ref{fig:a}, \ref{fig:c}, and \ref{fig:e}). `Centralized setup (no fairness)' are the unfairness measures without any bias mitigation, and `Centralized setup (w/fairness)' are the unfairness measures when applying the post-processing technique from Sec.~\ref{SEC:prelimsFair} in a centralized setup without any privacy protection. We ideally need privacy-preserving techniques that result in the unfairness metrics close to the `Centralized setup (w/fairness)'. Our results show that our method can still improve fairness even with the addition of our privacy-preserving techniques. Specifically, our method preserves users' privacy at every step of the process, both when users transfer $[\![\vectorbold{\hat{w}}^*]\!]$ and $[\![\vectorbold{\hat{r}}]\!]$ to the servers, and when the servers transfer $[\![\boldsymbol{\xi}]\!]$ to the users. We observe a trade-off between privacy and unfairness in our results, where a decrease in the privacy budget (\textit{i.e.}, more privacy) imposes a higher cost to the fairness, which is in line with the literature. This is because the amount of noise added perturbs the values of the unfairness measure, which consequently affects how well the ILP can compute rerankings when compared to the centralized setup without differential privacy. We note that our solution is able to preserve input privacy even with higher $\epsilon$.

\subsubsection{Utility vs.~Privacy Trade-offs }
We show the impact on reranking quality when introducing privacy for the users in Figure \ref{fig:results} (right column; Figures \ref{fig:b}, \ref{fig:d}, and \ref{fig:f}). $\textit{NDCG}=1$ represents the upper boundary of NDCG and indicates no change in the ordering of the relevance scores of the items compared to the original ranking. The dotted line at $\textit{NDCG}=0.8$ represents the quality constraint $\theta$ set in the ILP. Our results show that the quality of the rerankings are always maintained in the set boundary, $0.8 \leq \textit{NDCG} \leq 1$, irrespective of the amount of noise added to preserve privacy. Our study shows that using our approach, the privacy of users can be preserved without losing utility beyond the threshold $\theta$ initially set in the ILP.\\

\noindent
The resulting privacy-fairness trade-offs stem from preserving the output privacy. The utility-fairness trade-offs are due to the bias mitigation techniques and with and without privacy.

\subsubsection{Runtime }
All experiments were performed on a 2.6 GHz 6-Core Intel Core i7 with 16 GB RAM.
It takes about 0.67 seconds (averaged over $L=6,040$ users) to rerank a user's ranking for $n=100$ in a centralized setup. The additional cost in runtime due to added privacy is less than 5 seconds per client for a 2PC passive security setting with mixed circuits \cite{escudero2020improved}. These runtimes vary across different security settings. With a 3PC passive security setting \cite{araki2016high}, 
this additional runtime can be reduced to less than 1 second. We believe that this increased cost in runtime to rerank a user's ranking is worth the gain in privacy. We note that the MPC schemes are normally divided into two phases: the offline phase and the online phase, and we have reported runtimes for both. The offline phase performs computations independent of the data and thus can be carried out prior to the users' rerankings. By doing so, the responsiveness of our approach can be further improved to make our approach feasible in practice and near real-time. 

\section{Conclusion \& Future Work}
We presented the novel idea of promoting producer (item) fairness while preserving the privacy of the consumers (users) in a recommendation ecosystem using post-processing techniques. We proposed an approach in which users work in tandem with secure multi-party computation (MPC) servers to rerank items, taking into account both relevance scores and attention weights. The MPC servers receive user data in encrypted form only (secret shares), and perform all computations over this data while it stays encrypted. Furthermore, whenever the MPC servers need to release aggregated information to a user, they perturb it with noise to provide differential privacy (DP) guarantees, thereby avoiding leakage of the data of any user to any other user.

We demonstrated that the incorporation of our privacy-preserving approach results in unfairness mitigation without additional cost to utility, through comparison to the centralized approach in which all users disclose their data to a central server. 
Our approach can be extended to other bias mitigation techniques and various other notions of fairness in rankings. We believe our work promotes research possibilities at the intersection of privacy and fairness in recommender systems, while also encouraging development of techniques for end-to-end privacy-preserving and fairness promoting pipelines for both producers and consumers in multi-stakeholder recommender systems.

\subsubsection{Acknowledgements }
This project was partially funded by the Canadian Institute for Advanced Research (CIFAR).
\bibliography{references}

\begin{thebibliography}{10}
\providecommand{\url}[1]{\texttt{#1}}
\providecommand{\urlprefix}{URL }
\providecommand{\doi}[1]{https://doi.org/#1}

\bibitem{ammad2019federated}
Ammad-Ud-Din, M., Ivannikova, E., Khan, S.A., Oyomno, W., Fu, Q., Tan, K.E.,
  Flanagan, A.: Federated collaborative filtering for privacy-preserving
  personalized recommendation system. arXiv preprint arXiv:1901.09888  (2019)

\bibitem{araki2016high}
Araki, T., Furukawa, J., Lindell, Y., Nof, A., Ohara, K.: High-throughput
  semi-honest secure three-party computation with an honest majority. In: ACM
  SIGSAC Conference on Computer and Communications Security. pp. 805--817
  (2016)

\bibitem{biega2018equity}
Biega, A.J., Gummadi, K.P., Weikum, G.: Equity of attention: Amortizing
  individual fairness in rankings. In: 41st International ACM SIGIR Conference
  on Research \& Development in Information Retrieval. pp. 405--414 (2018)

\bibitem{canny2002collaborative}
Canny, J.: Collaborative filtering with privacy. In: IEEE Symposium on Security
  and Privacy. pp. 45--57 (2002)

\bibitem{celis2017ranking}
Celis, L.E., Straszak, D., Vishnoi, N.K.: Ranking with fairness constraints.
  In: 45th International Colloquium on Automata, Languages, and Programming
  (ICALP 2018). Leibniz International Proceedings in Informatics (LIPIcs),
  vol.~107, pp. 28:1--28:15 (2018)

\bibitem{chai2020secure}
Chai, D., Wang, L., Chen, K., Yang, Q.: Secure federated matrix factorization.
  IEEE Intelligent Systems  \textbf{36}(5),  11--20 (2020)

\bibitem{damgard}
Cramer, R., Damgard, I., Nielsen, J.: Secure Multiparty Computation and Secret
  Sharing. Cambridge University Press Print, New York (2015)

\bibitem{dehghani2017share}
Dehghani, M., Azarbonyad, H., Kamps, J., de~Rijke, M.: Share your model instead
  of your data: Privacy preserving mimic learning for ranking. arXiv preprint
  arXiv:1707.07605  (2017)

\bibitem{dwork2014algorithmic}
Dwork, C., Roth, A.: The algorithmic foundations of differential privacy.
  Foundations and Trends in Theoretical Computer Science  \textbf{9}(3-4),
  211--407 (2014)

\bibitem{escudero2020improved}
Escudero, D., Ghosh, S., Keller, M., Rachuri, R., Scholl, P.: Improved
  primitives for {MPC} over mixed arithmetic-binary circuits. In: Annual
  International Cryptology Conference. pp. 823--852. Springer (2020)

\bibitem{ge2021privitem2vec}
Ge, Z., Liu, X., Li, Q., Li, Y., Guo, D.: {PrivItem2Vec:} a privacy-preserving
  algorithm for {top-N} recommendation. International Journal of Distributed
  Sensor Networks  \textbf{17}(12) (2021)

\bibitem{joachims2007evaluating}
Joachims, T., Granka, L., Pan, B., Hembrooke, H., Radlinski, F., Gay, G.:
  Evaluating the accuracy of implicit feedback from clicks and query
  reformulations in web search. ACM Transactions on Information Systems (TOIS)
  \textbf{25}(2),  7--es (2007)

\bibitem{joachims2007search}
Joachims, T., Radlinski, F.: Search engines that learn from implicit feedback.
  Computer  \textbf{40}(8),  34--40 (2007)

\bibitem{kay2015unequal}
Kay, M., Matuszek, C., Munson, S.A.: Unequal representation and gender
  stereotypes in image search results for occupations. In: Proceedings of the
  33rd Annual ACM Conference on Human Factors in Computing Systems. pp.
  3819--3828 (2015)

\bibitem{mpspdz}
Keller, M.: {MP-SPDZ}: A versatile framework for multi-party computation. In:
  Proceedings of the 2020 ACM SIGSAC Conference on Computer and Communications
  Security. pp. 1575--1590 (2020)

\bibitem{kharitonov2019federated}
Kharitonov, E.: Federated online learning to rank with evolution strategies.
  In: Proceedings of the 12th ACM International Conference on Web Search and
  Data Mining. pp. 249--257 (2019)

\bibitem{liu2019real}
Liu, Y., Ge, K., Zhang, X., Lin, L.: Real-time attention based look-alike model
  for recommender system. In: Proceedings of the 25th ACM SIGKDD International
  Conference on Knowledge Discovery \& Data Mining. pp. 2765--2773 (2019)

\bibitem{mehrotra2018towards}
Mehrotra, R., McInerney, J., Bouchard, H., Lalmas, M., Diaz, F.: Towards a fair
  marketplace: Counterfactual evaluation of the trade-off between relevance,
  fairness \& satisfaction in recommendation systems. In: Proceedings of the
  27th ACM International Conference on Information and Knowledge Management.
  pp. 2243--2251 (2018)

\bibitem{morik2020controlling}
Morik, M., Singh, A., Hong, J., Joachims, T.: Controlling fairness and bias in
  dynamic learning-to-rank. In: Proceedings of the 43rd international ACM SIGIR
  Conference on Research \& Development in Information Retrieval. pp. 429--438
  (2020)

\bibitem{pentyala2022privfairfl}
Pentyala, S., Neophytou, N., Nascimento, A., De~Cock, M., Farnadi, G.:
  Privacy-preserving group fairness in cross-device federated learning. In:
  Proceedings of NeurIPS workshop on Algorithmic Fairness through the Lens of
  Causality and Privacy, Proceedings of Machine Learning Research (2023)

\bibitem{resheff2018privacy}
Resheff, Y.S., Elazar, Y., Shahar, M., Shalom, O.S.: Privacy and fairness in
  recommender systems via adversarial training of user representations. arXiv
  preprint arXiv:1807.03521  (2018)

\bibitem{sapiezynski2019quantifying}
Sapiezynski, P., Zeng, W., E~Robertson, R., Mislove, A., Wilson, C.:
  Quantifying the impact of user attention on fair group representation in
  ranked lists. In: Companion Proceedings of the 2019 World Wide Web
  Conference. pp. 553--562 (2019)

\bibitem{sato2022private}
Sato, R.: Private recommender systems: How can users build their own fair
  recommender systems without log data? In: Proceedings of the 2022 SIAM
  International Conference on Data Mining (SDM). pp. 549--557. SIAM (2022)

\bibitem{singh2018fairness}
Singh, A., Joachims, T.: Fairness of exposure in rankings. In: Proceedings of
  the 24th ACM SIGKDD International Conference on Knowledge Discovery \& Data
  Mining. pp. 2219--2228 (2018)

\bibitem{singh2019policy}
Singh, A., Joachims, T.: Policy learning for fairness in ranking. Advances in
  Neural Information Processing Systems  \textbf{32} (2019)

\bibitem{wang2021effective}
Wang, S., Liu, B., Zhuang, S., Zuccon, G.: Effective and privacy-preserving
  federated online learning to rank. In: Proceedings of the 2021 ACM SIGIR
  International Conference on Theory of Information Retrieval. pp. 3--12 (2021)

\bibitem{wang2021federated}
Wang, S., Zhuang, S., Zuccon, G.: Federated online learning to rank with
  evolution strategies: a reproducibility study. In: European Conference on
  Information Retrieval. pp. 134--149. Springer (2021)

\bibitem{wu2021fedgnn}
Wu, C., Wu, F., Cao, Y., Huang, Y., Xie, X.: {FedGNN:} federated graph neural
  network for privacy-preserving recommendation. arXiv preprint
  arXiv:2102.04925  (2021)

\bibitem{yang2018privacy}
Yang, D., Qu, B., Cudr{\'e}-Mauroux, P.: Privacy-preserving social media data
  publishing for personalized ranking-based recommendation. IEEE Transactions
  on Knowledge and Data Engineering  \textbf{31}(3),  507--520 (2018)

\bibitem{yang2017measuring}
Yang, K., Stoyanovich, J.: Measuring fairness in ranked outputs. In:
  Proceedings of the 29th International Conference on Scientific and
  Statistical Database Management. pp.~1--6 (2017)

\bibitem{zehlike2017fa}
Zehlike, M., Bonchi, F., Castillo, C., Hajian, S., Megahed, M., Baeza-Yates,
  R.: Fa* ir: A fair top-k ranking algorithm. In: Proceedings of the 2017 ACM
  on Conference on Information and Knowledge Management. pp. 1569--1578 (2017)

\end{thebibliography}
\bibliographystyle{splncs04}
\end{document}